\documentclass[runningheads]{svmult}

\usepackage{makeidx}   
\usepackage{graphicx}  
\usepackage{subeqnar}  
\usepackage{multicol}  
\usepackage{physprbb}  
\makeindex             

\newcommand{\greeksym}[1]{{\usefont{U}{psy}{m}{n}#1}}
\newcommand{\umu}{\mbox{\greeksym{m}}}

\def\mbox{\hbox}           

\def\deg{\ifmmode ^\circ                
         \else $^\circ$
         \fi
         \hskip -0.1truecm}
\def\degd#1.#2{                         
               \ifmmode {#1^{\hskip 0.05em\circ}\hskip-0.42em.\hskip0.08em#2}
               \else {#1$^{\hskip 0.05em\circ}\hskip-0.42em.\hskip0.08em$#2}
               \fi
              }
\def\mind#1.#2{                         
               \ifmmode {#1^{\hskip 0.05em\prime}\hskip-0.35em.\hskip0.05em#2}
               \else {#1$^{\hskip 0.05em\prime}\hskip-0.35em.\hskip0.05em$#2}
               \fi
              }
\def\secd#1.#2{                         
               \ifmmode {#1^{\prime\prime}\hskip-0.46em.\hskip0.12em#2}
               \else {#1$^{\prime\prime}\hskip-0.46em.\hskip0.12em$#2}
               \fi
              }
\def\timsecd#1.#2{                      
                  \ifmmode {#1^{\rm s}\hskip-0.39em.\hskip0.08em#2}
                  \else {$#1^{\rm s}\hskip-0.39em.\hskip0.08em#2$}
                  \fi
                 }
\def\hms#1h#2m#3s{                      
                  \relax
                  \ifmmode #1^{\rm h}\,#2^{\rm m}\,#3^{\rm s}
                  \else \hbox{$#1^{\rm h}\,#2^{\rm m}\,#3^{\rm s}$}
                  \fi
                 }
\def\dms#1d#2m#3s{                      
                  \relax
                  \ifmmode #1^\circ\,#2^{\prime}\,#3^{\prime\prime}
                  \else \hbox{$#1^\circ\,#2^{\prime}\,#3^{\prime\prime}$}
                  \fi
                 }
\def\dmsd#1d#2m#3.#4s{                  
                      \relax
                      \ifmmode #1^\circ\,#2^{\prime}\,#3^{\prime\prime}
                               \hskip-0.46em.\hskip0.12em#4
                      \else \hbox{$#1^\circ\,#2^{\prime}\,#3^{\prime\prime}
                            \hskip-0.46em.\hskip0.12em#4$}
                      \fi
                     }
\def\hm#1h#2m{                          
              \relax
              \ifmmode #1^{rm h}\,#2^{\rm m}
              \else \hbox{$#1^{\rm h}\,#2^{\rm m}$}
              \fi
             }
\def\dm#1d#2m{                          
              \relax
              \ifmmode #1^\circ\,#2^{\prime}
              \else \hbox{$#1^\circ\,#2^{\prime}$}
              \fi
             }
\def\hmsd#1h#2m#3.#4s{                  
                      \relax
                      \ifmmode #1^{\rm h}\,#2^{\rm m}\,#3^{\rm s}
                               \hskip-0.39em.\hskip0.08em#4
                      \else \hbox{$#1^{\rm h}\,#2^{\rm m}\,#3^{\rm s}
                            \hskip-0.39em.\hskip0.08em#4$}
                      \fi
                     }
\def\hmd#1h#2.#3m{                  
                  \relax
                  \ifmmode #1^{\rm h}\,#2^{\rm m}
                           \hskip-0.55em.\hskip0.22em#3
                  \else \hbox{$#1^{\rm h}\,#2^{\rm m}
                        \hskip-0.55em.\hskip0.22em#3$}
                  \fi
                 }
\def\mg{\relax                          
        \ifmmode ^{\rm m}
        \else $^{\rm m}$
        \fi
       }
\def\mgd#1.#2{                          
              \relax
              \ifmmode #1^{\rm m}
                       \hskip-0.55em.\hskip0.22em#2
              \else \hbox{#1$^{\rm m}
                    \hskip-0.55em.\hskip0.22em$#2}
              \fi
             }

\def\la{\mathrel{\hbox{\rlap{\hbox{\lower4pt\hbox{$\sim$}}}\hbox{$<$}}}}
\def\ga{\mathrel{\hbox{\rlap{\hbox{\lower4pt\hbox{$\sim$}}}\hbox{$>$}}}}

\def\unitspace{\;}                      

\def\un#1{\ifmmode \unitspace\mbox{\rm #1} 
          \else $\unitspace$#1
          \fi}
\def\pun#1#2{\ifmmode \unitspace\mbox{\rm #1}^{#2} 
             \else $\unitspace$#1$^{#2}$
             \fi}

\def\kms{\un{km}\pun{s}{-1}}          
\def\Lsun{\ifmmode \un{L}_{\odot}     
          \else $\un{L}_{\odot}$
          \fi}
\def\Msun{\ifmmode \un{M}_{\odot}     
          \else $\un{M}_{\odot}$
          \fi}
\def\mum{\ifmmode \unitspace\umu\mbox{\rm m} 
         \else $\unitspace\umu$m
         \fi}
\def\pyr{\pun{yr}{-1}}                
\def\sqarcsec{\ifmmode \unitspace\Box''    
              \else $\unitspace\Box''$     
              \fi} 

\def\Bp{\relax                            
        \ifmmode B_{||}                   
        \else $B_{||}$
        \fi}
\def\Bt{\relax                            
        \ifmmode B\!_{\perp}              
        \else $B\!_{\perp}$               
        \fi}
\def\Gcr{\relax                           
         \ifmmode \Gamma\!_{\rm cr}       
         \else $\Gamma\!_{\rm cr}$
         \fi}
\def\ICII{\relax                          
          \ifmmode I_{[\CII]}             
          \else $I_{[\CII]}$
          \fi}
\def\LHtwo{\relax                                 
           \ifmmode L_{\mbox{\rm\scriptsize H}_2} 
           \else $L_{\mbox{\rm\scriptsize H}_2}$  
           \fi}
\def\LLya{\relax                          
          \ifmmode L_{{\rm Ly}\,\alpha}   
          \else $L_{{\rm Ly}\,\alpha}$
          \fi}
\def\MHtwo{\relax                                 
           \ifmmode M_{\mbox{\rm\scriptsize H}_2} 
           \else $M_{\mbox{\rm\scriptsize H}_2}$  
           \fi}
\def\MHtwodot{\relax                                       
              \ifmmode \dot{M}_{\mbox{\rm\scriptsize H}_2} 
              \else $\dot{M}_{\mbox{\rm\scriptsize H}_2}$  
              \fi}                                         
\def\Mstardot{\relax                      
              \ifmmode \dot{M}_{\ast}     
              \else $\dot{M}_{\ast}$      
              \fi}
\def\nHI{\relax                                      
         \ifmmode n_{\mbox{\scriptsize\rm H\,\sc I}} 
         \else $n_{\mbox{\scriptsize\rm H\,\sc I}}$
         \fi}
\def\nHtwo{\relax                                
           \ifmmode n_{{\mbox{\scriptsize H}}_2} 
           \else $n_{{\mbox{\scriptsize H}}_2}$  
           \fi}
\def\rhostardot{\relax                         
                \ifmmode \dot{\rho}_{\ast}     
                \else $\dot{\rho}_{\ast}$      
                \fi}
\def\rhoZdot{\relax                          
             \ifmmode \dot{\rho}_{\rm Z}     
             \else $\dot{\rho}_{\rm Z}$      
             \fi}

\def\sou#1#2{\relax                       
             \ifmmode {\rm #1}\,{\rm #2}  
             \else #1$\,$#2
             \fi}

\def\NGC#1{\sou{NGC}{#1}}                

\def\Arp#1{\sou{Arp}{#1}}                

\def\qu#1#2{\relax                          
            \ifmmode #1_{\rm #2}            
            \else $#1_{\rm #2}$
            \fi}

\def\CO#1{\ifnum#1=0                    
           \ifmmode \mbox{\rm CO}
           \else {\rm CO}
           \fi
          \else
           \ifnum#1<15
            \ifmmode ^{#1}\mbox{\rm CO}
            \else $^{#1}${\rm CO}
            \fi
           \else
            \ifmmode \mbox{\rm C}^{#1}\mbox{\rm O}
            \else {\rm C}$^{#1}${\rm O}
            \fi
           \fi
          \fi}

\def\COp{\ifmmode \mbox{\rm CO}^+           
         \else {\rm CO}$^+$                 
         \fi}

\def\CS#1{\ifnum#1=0                    
           \ifmmode \mbox{\rm CS}
           \else {\rm CS}
           \fi
          \else
           \ifnum#1<15
            \ifmmode ^{#1}\mbox{\rm CS}
            \else $^{#1}${\rm CS}
            \fi
           \else
            \ifmmode \mbox{\rm C}^{#1}\mbox{\rm S}
            \else {\rm C}$^{#1}${\rm S}
            \fi
           \fi
          \fi}

\def\HCOp{\ifmmode \mbox{\rm HCO}^+          
          \else {\rm HCO}$^+$                
          \fi}
\def\Hthreep{\ifmmode \mbox{\rm H}_3^+         
             \else {\rm H}$_3^+$               
             \fi}

\def\Htwo{\ifmmode \mbox{\rm H}_2              
          \else {\rm H}$_2$                    
          \fi}

\def\HtwoO{\ifmmode \mbox{\rm H}_2\mbox{\rm O} 
           \else {\rm H}$_2${\rm O}            
           \fi}

\def\ion#1#2{\ifmmode \mbox{{\rm #1}}\,\mbox{{\sc #2}} 
        \else {\rm #1}$\,${\sc #2}
        \fi}

\def\HII{\ion{H}{ii}}

\def\rec#1#2{\if#2a                            
              \ifmmode \mbox{{\rm #1}}\alpha   
              \else {\rm #1}$\alpha$
              \fi
             \fi
             \if#2b
              \ifmmode \mbox{{\rm #1}}\beta
              \else {\rm #1}$\beta$
              \fi
             \fi
             \if#2g
              \ifmmode \mbox{{\rm #1}}\gamma
              \else {\rm #1}$\gamma$
              \fi
             \fi}

\def\Ha{\rec{H}{a}}                            
\def\Bra{\rec{Br}{a}}                          
\def\Brg{\rec{Br}{g}}                          

\newcommand{\figref}[1]{Fig.~\protect\ref{#1}}

\newcommand{\eqref}[1]{Eq.~$\left(\protect\ref{#1}\right)$}

\newcommand{\twoeqsref}[2]{$\left(\protect\ref{#1}\right)$--$\left(\protect\ref{#2}\right)$}
\newcommand{\secref}[1]{Sect.~\protect\ref{#1}}

\begin{document}
\title*{Starbursts In Ultraluminous Infrared Galaxies - Fueling And Properties}
%
%
%
%
\titlerunning{Starbursts in ultraluminous infrared galaxies}
%
\author{Paul P. van der Werf}
\authorrunning{Paul van der Werf}
%
%
\institute{Leiden Observatory, P.O.~Box~9513,\\ 
           NL~2300~RA Leiden, The Netherlands}

\maketitle              

\makeatletter
\renewcommand{\@makefnmark}{\mbox{\ }}
\makeatother

\renewcommand{\thefootnote}{}

\footnote{to appear in {\it
``Starburst galaxies: near and far''}, eds.\ D.\ Lutz \& L.\ J.\ Tacconi,
(Springer)}       

\begin{abstract}
The properties of starbursts in ultraluminous infrared galaxies are
discussed, with particular emphasis on the fueling, the amount of
extinction and the intrinsic properties of the nuclear
starbursts.
It is shown by the example of $\NGC{6240}$ 
that the $\Htwo$ vibrational lines can be used to measure
the rate of gas inflow into the potential well, which is sufficient to
fuel a nuclear starburst of the intensity required to account for the
far-infrared emission.
It is shown that in $\Arp{220}$ the faintness of all tracers of ionized gas
can be accounted for by Lyman continuum absorption by
dust within the ionized regions, combined with significant (but not
extreme) extinction; there is no reason to invoke the presence of 
extreme extinction,
an old starburst, or
an additional non-stellar power source in $\Arp{220}$.
\end{abstract}

\begin{figure}[t]
\begin{center}
\includegraphics[width=.65\textwidth,angle=270]{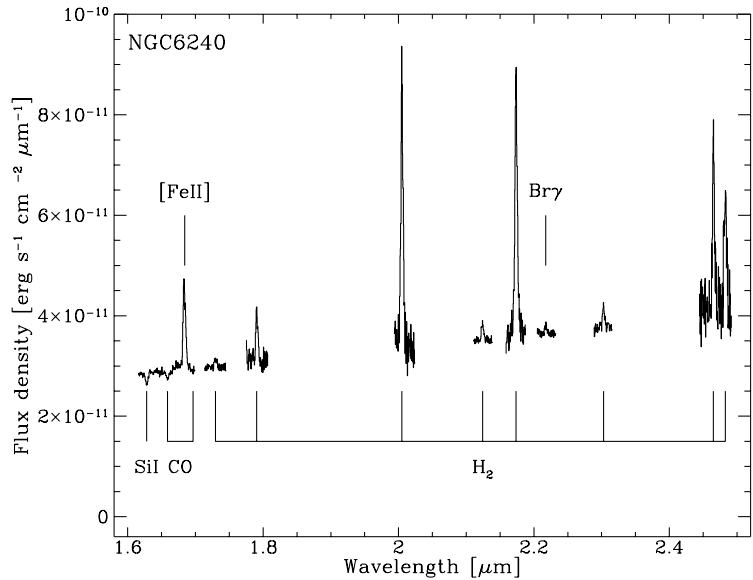}
\end{center}
\caption[]{Spectra of $\NGC{6240}$ in selected parts of the near-infrared $H$
and $K$ bands, integrated over a $\secd 4.4$ region~\cite{VanDerWerfetal01}}
\label{fig.NGC6240spec}
\end{figure}

\section{Introduction}
\label{sec.intro}

Are ultraluminous infrared galaxies (ULIGs) powered by intense bursts
of star formation or is an additional source of energy, such as an
active galactic nucleus (AGN) required? In order to address this
question, it is instructive to compare the near-infrared
spectra of ULIGs and lower luminosity starburst galaxies. In starburst
galaxies of low or moderate luminosity, the dominant emission line is
the $\Brg$ line, underlining the importance of massive young stars in
the energetics of these objects. 
The brightest $\Htwo$ rovibrational line, the $\Htwo$
$v=1{\to}0$ S(1) line, is typically
fainter~\cite{MoorwoodOliva88,MoorwoodOliva90,Puxleyetal88,Puxleyetal90,Vanzietal96,VanziRieke97}.
In contrast, in ULIGs the $\Htwo$ $v=1{\to}0$ S(1) line is
significantly brighter than the $\Brg$ line, which is often not even
detected~\cite{Goldaderetal95,Murphyetal01}. This behaviour
shows that ULIGs are not just scaled-up starburst galaxies.
An extreme case of is presented by $\NGC{6240}$ (\figref{fig.NGC6240spec}),
where the $\Htwo$ $v=1{\to}0$ S(1) line is 40 times brighter than
$\Brg$. The low ratio of $\Brg$ luminosity to FIR luminosity in ULIGs
has been used to argue against star formation as the power
source of ULIGs, for instance in the nearby ULIG $\sou{Arp}{220}$, 
where a starburst with the low $\Brg$
luminosity observed can account for at most 10\% of the
bolometric luminosity of the galaxy~\cite{Armusetal95,Scovilleetal97}. However,
spectroscopy of ULIGs at longer wavelengths with the Infrared Space
Observatory (ISO)
revealed bright emission lines from powerful starbursts, 
which are the
dominant power source in most of the objects studied, but are obscured
at shorter wavelengths~\cite{Genzeletal98}. Motivated by these results,
in this paper the physical processes revealed by the $K$-band spectra
of ULIGs are reexamined in the context of the starburst scenario. In
\secref{sec.NGC6240} the origin of the $\Htwo$ line emission is
discussed, while \secref{sec.Arp220} addresses the faintness of $\Brg$.

\begin{figure}[t]
\begin{center}
\includegraphics[width=.5\textwidth,angle=270]{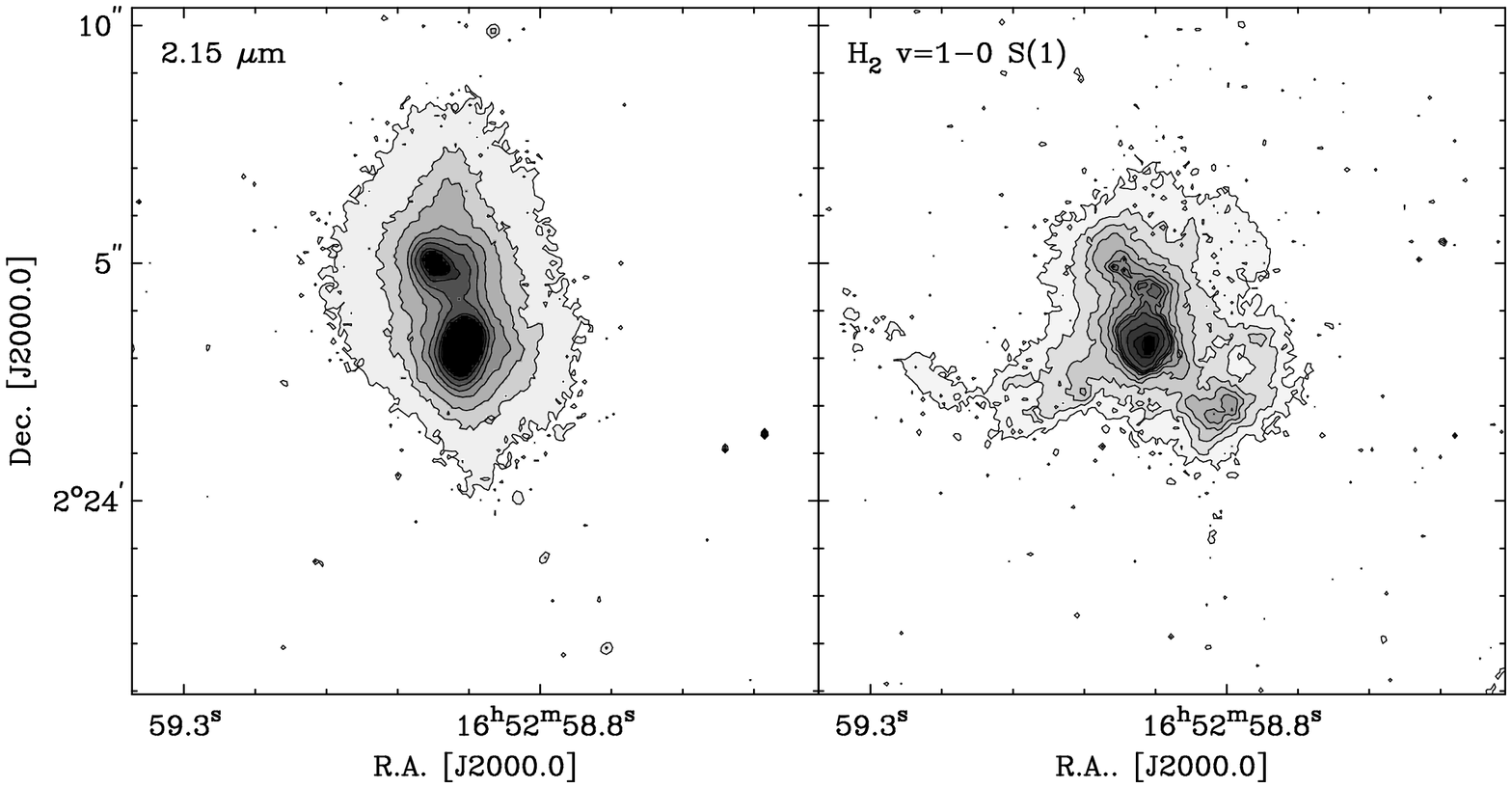}
\end{center}
\caption[]{High resolution imaging of $\NGC{6240}$ in the $\Htwo$
$v=1{\to}0$ S(1) line and the $2.15\mum$ continuum with 
NICMOS/HST~\cite{VanDerWerfetal01}}
\label{fig.NGC6240NICMOS}
\end{figure} 

\section{Origin of the H$_2$ emission in ULIGs}
\label{sec.NGC6240}

The most extreme 
vibrational $\Htwo$ emission is found in the nearby luminous merger
$\NGC{6240}$: $7\cdot10^7\Lsun$ is emitted 
in the $\Htwo$ $v=1{\to}0$ S(1) line alone (for
$H_0=75\kms\pun{Mpc}{-1}$ and with no correction for extinction). 
This line contains 0.012\% of the
bolometric luminosity of $\NGC{6240}$, which is considerably higher 
than any other galaxy \cite{VanDerWerfetal93}. Together the
vibrational lines may account for 0.1\% of the total bolometric luminosity.

Imaging of the $\Htwo$ $v=1{\to}0$ S(1) emission from $\NGC{6240}$
has shown that the $\Htwo$ emission peaks \emph{between\/} the two
remnant nuclei of the merging system \cite{VanDerWerfetal93}. 
This morphology provides a
unique constraint on the excititation mechanism, since it argues
against any scenario where the excitation is dominated by the stellar
component (e.g., UV-pumping, excitation by shocks or X-rays from
supernova remnants). Instead, the favoured excitation mechanism is
slow shocks in the nuclear gas component, which, as shown by 
high resolution interferometry in the CO 
$J=2{\to}1$ line \cite{Tacconietal99},
also peaks between the nuclei of $\NGC{6240}$.

What is the role of these shocks? In the shocks mechanical
energy is dissipated and radiated away, mostly in spectral lines
(principally $\Htwo$, CO, $\HtwoO$ and [$\ion{O}{i}$] lines). This
energy is radiated away at the expense of the
orbital energy of the molecular clouds in the central potential well. 
Consequently, the dissipation of mechanical energy by the shocks will
give rise to an infall of molecular gas to the centre of the potential 
well. Therefore,
\emph{the $\Htwo$ vibrational lines measure the rate of infall of
molecular gas\/} into the central potential well. 
This conclusion can be quantified by writing
\begin{equation}
\qu{L}{rad}=\qu{L}{dis},
\label{eq.L}
\end{equation}
where $\qu{L}{rad}$ is the total luminosity radiated by the shocks and
$\qu{L}{dis}$ the dissipation rate of mechanical energy, giving rise
to a molecular gas infall rate $\dot{M}_{{\rm H}_2}$ given by
\begin{equation}
\qu{L}{dis}={1\over2}\dot{M}_{{\rm H}_2}v^2,
\label{eq.MH2dot}
\end{equation}
where $v$ is the circular
orbital velocity at the position where the shock occurs.

Using a $K$-band extinction of $\mgd 0.15$ \cite{VanDerWerfetal93},
the total luminosity of $\Htwo$ vibrational lines from $\NGC{6240}$
becomes $7.2\cdot10^8\Lsun$; inclusion of the purely rotational lines
observed with ISO approximately doubles this number, so that
$\qu{L}{rad}=1.5\cdot10^9\Lsun$. 

In order to use this number to estimate $\dot{M}_{{\rm H}_2}$, it is
necessary to establish more accurately the fraction of the $\Htwo$
emission that is due to infalling gas. Observations with NICMOS on the 
Hubble Space Telescope (HST) provide the required information
(\figref{fig.NGC6240NICMOS})~\cite{VanDerWerfetal01}.
The NICMOS image shows that the emission
consists of a number of tails (presumably related to the superwind
also observed in $\Ha$ emission), and concentrations assocated with
the two nuclei, and a further concentration approximately (but not
precisely) between the two nuclei. The relative brightness of the
$\Htwo$ emission from the southern nucleus is deceptive, since this
nucleus is much better centred in the filter that was used for these
observations than the other emission components, in particular the
northern nucleus. Taking this effect into account, it is found that
32\% of the total $\Htwo$ flux is associated with the southern
nucleus, 16\% is associated with the northern nucleus, and 12\% with
the component between the two nuclei, the remaining 40\% being
associated with extended emission. Using inclination-corrected
circular velocities of $270$ and $360\,\kms$ 
for the southern and northern nucleus respectively~\cite{Teczaetal00}, 
and of $280\kms$
for the central component \cite{Tacconietal99}, the mass infall rates 
derived using \twoeqsref{eq.L}{eq.MH2dot} are $80\Msun\pun{yr}{-1}$
for the southern nucleus, $22\Msun\pun{yr}{-1}$ for the northern
nucleus and $28\Msun\pun{yr}{-1}$ for the central component.

The derived
molecular gas inflow rate to the two nuclei is remarkably close to the 
mass consumption rate by star formation of approximately $60\Msun\pyr$, 
indicating that the $\Htwo$ emission from the nuclei directly measures 
the fueling of the starbursts in these regions. This analysis shows
that the central regions of $\NGC{6240}$ are being fueled at a rate
sufficient to maintain starburst activity at the level required to
account for the FIR luminosity.

\section{Dusty, compact starbursts in ULIGs}
\label{sec.Arp220}

Accepting the starburst model, the faintness of $\Brg$ and other
recombination lines remains to be addressed. As pointed out in
\secref{sec.intro}, it is evident that
selective extinction towards the regions of most
recent massive star formation plays a significant role in suppressing
the $\Brg$ emission. Can this effect be quantified? 

\begin{figure}[t]
\begin{center}
\includegraphics[width=.5\textwidth]{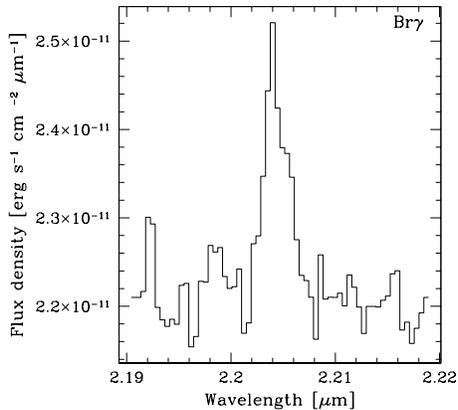}
\end{center}
\caption[]{Integrated spectrum of $\Brg$ emission from 
$\Arp{220}$~\cite{VanDerWerfIsrael01}}
\label{fig.Arp220Brg}
\end{figure}

The nearby ULIG $\Arp{220}$ has been studied in detail with the ISO
satelite. Based on an upper limit on the ratio of the [$\ion{S}{iii}$]
18 and $33\mum$ lines, and a high $\Bra$ over $\Brg$ flux ratio, an
extinction $A_V=50\pm10\mg$, located purely in an absorbing foreground
screen was proposed~\cite{Sturmetal96}. However, the supporting
arguments have now weakened significantly. In the first place, a
better understanding of the calibration of the [$\ion{S}{iii}$] $33\mum$
spectrum has made the upper limit on the [$\ion{S}{iii}$] 18 to $33\mum$
flux ratio less strict by approximately a factor of
two~\cite{Genzeletal98}. Secondly, the $\Bra$ flux from
ISO~\cite{Sturmetal96} is almost certainly an overestimate: the $\Bra$
line
displays, on top of a higly structured baseline, a double-peaked
structure, which is absent in any other line (including
long-wavelength lines such as the well-detected [$\ion{S}{iii}$]
$33\mum$ line). The velocity difference between the two peaks in
$\Bra$ is approximately $600\kms$
and therefore cannot be attributed
to motion of the two nuclei of $\Arp{220}$, which have a radial
velocity difference of approximately $200\kms$~\cite{Scovilleetal97}.
The same velocity difference of $200\kms$ is found in long-slit
$\Brg$ spectra of the $\Arp{220}$ nuclei~\cite{Larkinetal95}.
An integrated high-resolution $\Brg$ spectrum
(\figref{fig.Arp220Brg}) shows no trace of a double-peaked
structure, indicating that the high $\Bra$ flux found with ISO is most
likely dominated by structure in the spectral baseline. The extinction
derived from the
$\Bra$/$\Brg$ ratio should thus be used as an upper limit.
A further argument against an obscuring foreground screen
with $A_V\approx50\mg$ is furnished by the derived Lyman continuum
fluxes, which increase as
as shorter wavelength tracers are used~\cite{Genzeletal98}, a
behaviour suggesting that the extinction has been overestimated. 
Finally, a strong limit on the presence of an intense, highly
obscured, but otherwise normal starburst
follows from the upper limit to the free-free
emission at millimetre wavelengths~\cite{Scovilleetal97}. 
These results indicate that either a foreground screen with a lower visual
extinction, or a model with mixed emission and
absorption needs to be adopted. 

This result does however \emph{not\/} imply the presence of a strongly
aged starburst, or an additional source of power in $\Arp{220}$, since
the preceding analysis ignored the role of
Lyman continuum absorption
by dust within the ionized regions. If most of the ionizing radiation
is absorbed by dust grains rather than hydrogen atoms, a dust-bounded
(rather than hydrogen-bounded) nebula results, and all tracers of
ionized gas (recombination lines, fine-structure lines, free-free
emission) will be suppressed. If the $\HII$ regions in $\Arp{220}$ are
principally dust-bounded, the
observational properties of $\Arp{220}$ can be accounted for, even
with only moderate extinction.  
Since the dust would also absorb far-ultraviolet radiation longwards
of the Lyman limit, the formation of photon-dominated regions would
also be suppressed, and the thus the same mechanism can account for
the faintness of the $158\mum$ [$\ion{C}{ii}$] line in
$\Arp{220}$ and other ULIGs~\cite{Fischeretal99}. 

Is the $\Arp{220}$ starburst dominated by dust-bounded $\HII$ regions?
The \emph{average\/} molecular gas density in the
$\sim10^{10}\Msun$ nuclear
molecular complex in $\Arp{220}$ is 
$\nHtwo\sim2\cdot10^4\pun{cm}{-3}$~\cite{Scovilleetal97}. The strong
emission from high dipole moment molecules such as CS, HCO$^+$ and HCN 
argues for even higher densities: $\sim10^{10}\Msun$ of molecular
gas (i.e., \emph{all\/} of the gas in the nuclear complex) has a density
$\nHtwo\sim10^5\pun{cm}{-3}$~\cite{Solomonetal90}.
At such densities the ionized nebulae created by hot stars are 
\emph{compact\/} or \emph{ultracompact\/} $\HII$ regions, where 
50 to 99\% of the Lyman continuum is absorbed by
dust~\cite{WoodChurchwell89}. Observationally, hydrogen-bounded
and dust-bounded $\HII$ regions can be distinguished by
the quantity $R=\qu{L}{FIR}$/$\qu{L}{\mathrm Br\gamma}$: for a wide range of
parameters, $R<3570$ implies that the nebula is hydrogen-bounded, while 
$R>35700$ implies that the nebula is dust-bounded~\cite{Bottorffetal98}.
For $\Arp{220}$, the $\Brg$ luminosity of $1.3\cdot10^6\Lsun$ (from
the spectrum in \figref{fig.Arp220Brg}, with a distance of
$77\un{Mpc}$) implies $R=1.6\cdot10^5$ assuming an obscuring foreground
screen with
$A_V=20\mg$ (a model consistent with the results discussed
above). Even with a foreground extinction of $A_V=50\mg$ (which is
most likely an overestimate, as discussed above), a ratio $R=1\cdot10^4$
would result, so that even in that case 
the absorption of Lyman continuum radiation by dust would
play a significant role.
The star formation takes place in (ultra)compact $\HII$
regions, where
all of the usual tracers of ionized gas (recombination lines,
fine-structure lines, free-free emission) are \emph{quenched}, not
extincted. While this result significantly complicates the interpretation of 
diagnostics of massive star formation in ULIGs, it is save to conclude that
the properties of $\Arp{220}$ can be accounted
for by an intense, and significantly (but not extremely) obscured
starburst. There is no reason to invoke the presence of extreme
extinction, a strongly aged starburst, or
an additional power source in $\Arp{220}$.

\end{document}